\documentclass[]{raa}
\usepackage{graphicx,times}	
\usepackage{amsmath}	
\usepackage{amssymb}	

\newcommand{\teff}{$T_{\rm eff}$}
\newcommand{\logg}{log$g$}

\newcommand{\kms}{km\ s$^{-1}$}
\newcommand{\mgb}{$\rm Mg_{b}$}

\begin{document}
\title{Red Clump Stars from LAMOST II: the outer disc of the Milky Way}
\author{Junchen Wan\inst{1}\inst{2}, Chao Liu\inst{1}, Licai Deng\inst{1}, Yue Wu\inst{1}}
\institute{$^1$Key Laboratory of Optical Astronomy, National Astronomical Observatories, Chinese Academy of Sciences, 20A Datun Road, Beijing, 100012, China;  {junchenwan@bao.ac.cn}\\
$^2$University of the Chinese Academy of Sciences, Beijing 100049, China catalogue
}

\abstract{
We present stellar density maps of the Galactic outer disc with red clump stars from the LAMOST data. These samples are separated into younger (mean age $\sim2.7$\,Gyr) and older (mean age $\sim4.6$\,Gyr) populations so that they can trace the variation of the structures with ages in the range of the Galactocentric radius $R$ from 9 to 13.5\,kpc. We show that both the scale heights for the two populations increase with $R$ and display radial gradients of $48\pm6$ and $40\pm4$\,pc/kpc for the older and younger populations, respectively. This is evident that the flaring occurs in the thin disc populations with a wide range of ages. Moreover, the intensity of flaring seems not significantly related to the age of the thin disc populations. On the other hand, the scale lengths of the radial surface density profiles are $4.7\pm0.5$\,kpc for the younger and $3.4\pm0.2$\,kpc for the older population, meaning that the younger disc population is more radially extended than the older one. Although the fraction of the younger population mildly increases from 28\% at $R\sim9$ to about 35\% at $R\sim13$\,kpc, the older population is prominent with the fraction no less than 65\% in the outer disc. 
}

\authorrunning{J. C. Wan et al.}            
   \titlerunning{Red clump stars from the LAMOST data II}  

   \maketitle


\section{Introduction}\label{sect:introduction}
Although it is often very faint and does not contain many stars, the outskirt of galaxies reflects the nature of the formation and evolution of the galactic discs. Simulations show that the secular evolution of the discs can significantly alter the properties, from the age distribution to the stellar density profile, of the outer disc (Ro{\v s}kar et al.~\cite{roskar2008}, Debattista et al.~\cite{debattista2006}). Meanwhile, the discs are easier to be perturbed either by dynamical torques or by new infalling gas. As a consequence, warp can be induced in their outskirts (Shen \& Sellwood~\cite{shen2006},  Ro{\v s}kar et al.~\cite{roskar2010}). Minor mergers may also affect the outer discs and induce substructures in the stellar density as well as in the kinematics (G{\'o}mez et al.~\cite{gomez2013,gomez2016}).

However, observations of the outskirt of disc galaxies are not trivial. In principle, the surface brightness of outer discs is extremely faint and imaging of the outskirt regions often suffers from very low signal-to-noise ratio. Nevertheless, Pohlen \& Trujillo~(\cite{pohlen2006}) classified several hundreds external galaxies by their surface luminosity distributions and found that the outer discs show three different types of profiles, exponential, down-bending, and up-bending. Zheng et al.~(\cite{zheng2015}) also showed that the surface brightness profiles of disc galaxies are broken in bluer bands, while they show unbroken exponential in redder bands. The mechanisms leading to these differences are not quite clear so far. 

For the case of the Milky Way, since it allows for observations of individual stars, it takes a great advantage in the studies of the Galactic outer disc. L{\'o}pez-Corredoira et al.~(\cite{lopez02}) fitted the exponential density distribution models of disc by using 2MASS red clump stars and found that the Galactic outer disc is substantially flared and warped. They suggested that the flaring disc can be described with an exponentially increasing scale height with radius. Moreover, a puzzling substructure, the Monoceros ring, in the Galactic anti-centre direction has been unveiled by Newberg et al.~(\cite{newberg2002}) from the star count of turnoff stars. Momany et al.~(\cite{momany2006}) claimed that this is likely the effect of the warp and flare by using red clump stars and giant stars selected from 2MASS. G{\'o}mez et al.~(\cite{gomez2016}) analyzed the vertical structure in simulation data, and their results showed that the interaction among the satellites, Galactic halo and the disc can also induce such kind of feature. Furthermore, Xu et al.~(\cite{xu15}) discovered that the star counts in the north and south of the Galactic mid-plane of the outer disc are not symmetric but display wave-like oscillations. Recently, Liu et al.~(\cite{liu2017a}) found that the disc has no truncation but exponentially expands to 19\,kpc and then smoothly transition to the stellar halo.

Not only are the spatial structures quite complicated, the stellar kinematics in the outer disc also deviate from axisymmetry. Tian et al.~(\cite{tian16}) showed that the mean radial velocity is not zero but varies from $-6$ to $+7$\,\kms\ with Galactocentric radius from 7 to 14\,kpc with red clump stars from LAMOST data. Such an oscillated radial velocity may be associated with the perturbation induced by the rotating bar or spiral arms. Meanwhile, Liu, Tian, \& Wan~(\cite{liu2017b}) demonstrated that the stellar warp may also lead to vertical peculiar velocity in some directions, especially for young stars.

The nature of the outer disc can be learnt not only from the spatial structures and stellar kinematics, but also from the features of stellar populations. In general, chemical abundances and age can well characterise stellar populations. However, elemental abundances can only be available for high signal-to-noise and high spectral resolution data, which are rarely observed in the outer disc. On the other hand, the age of field stars is probably the hardest quantity to be determined. To avoid the non-trivial works of determining the age for individual stars, we turn to use some special tracers for stellar disc populations with different ages. 

In this work, we select red clump (RC) stars obtained from the LAMOST survey (Cui et al.~\cite{cui12}, Zhao et al.~\cite{zhao12}) as the tracers and compare the structural parameters between younger and older RC populations. The LAMOST survey has well covered the Galactic anti-centre region due to the special conditions of the site of the telescope (Deng et al.~\cite{deng12}, Yao et al.~\cite{yao12}). Thus, it can well sample the Galactic outer disc beyond the location of the Sun.

The paper is organized as follow. In Section~\ref{sect:data}, we outline the RC stars catalog identified from LAMOST data, purifying and separating the RC candidates samples and the methods to derive the stellar density profiles with spectroscopic data. In Section~\ref{sect:result}, we show the stellar density maps for the younger and older RC populations, respectively. The scale heights as functions of $R$ and the scale lengths for the two populations are determined. The fraction of the younger population varying with $R$ is also demonstrated in this section. In Section~\ref{sect:conclusion}, we discuss the influence from the interstellar extinction, warp and comparison between our results with others. In the same section, we briefly draw conclusions.

\section{Data}\label{sect:data}
Wan et al.~(\cite{wan2015}) (thereafter, Paper I) identified more than 100 thousands of RC candidates with [Fe/H]$>-0.6$\,dex from the LAMOST DR2 catalogue using the seismic calibrated surface gravities from Liu et al.~(\cite{liu2015}). We adopt a similar approach to identify more RC candidates from the LAMOST DR3 catalogue, which totally contains more than 4 million stellar spectra. 

Then, we apply the criteria suggested by Tian et al.~(\cite{tian16}) (thereafter, Paper II) to purify the RC candidates samples. Paper II drew the distribution of RC candidates in the [Fe/H] vs. \mgb\, plane, and adopted an empirical separation line to cut the RGB stars off (see their Figure.1). Although \mgb\, is an $\alpha$ element, it is more sensitive to \logg\, and \teff. Therefore, the [Fe/H] vs. \mgb\, should not significantly reflect the $\alpha$-abundance. It is also noted that, according to Hayden et al.~(\cite{hayden15}), the outer disc is dominated by low-$\alpha$ stars. In other words, the RC stars should not cover a large range of $\alpha$ abundances. Generally speaking, with the same [Fe/H] and \logg, RC stars have larger \teff\, than RGB stars. On the other hand, with the same [Fe/H] and \logg, RC stars have smaller \logg. Therefore, RC stars and RGB stars should be separate in the [Fe/H] vs. \mgb\, relation. In order to verify the separation line, Paper II  overlapped the cross-matched RC stars from Stello et al.~(\cite{stello13}), who used seismic period and frequency to accurately classify primary, secondary RC stars and RGB stars. Paper II found that about 97\% seismic-identified RC stars are classified by the empirical separation line and 94\% for RGB stars.

Paper II also suggested to separate the RC samples into younger and older populations from the effective temperature--surface gravity--metallicity space. Because the initial stellar mass of secondary RC stars ($\sim \rm 2M_{\sun}$) are larger than primary ones, secondary RC stars have non-degenerate helium-burning cores. The different evolution tracks for primary and secondary RC stars can be present at different positions in \teff-\logg\, plane. Paper II used PARSEC isochrones (Bressan et al.~\cite{bressan12}) to distinguish RC stars younger or older than 2\,Gyr in the \teff-\logg\, plane with different metallicities (see their Figure.2). According to Paper II (see their Figure.3), the younger RC population (hereafter RCyoung), which has a mean age of about 2.7\,Gyr (covering age from $\sim$1\,Gyr to 6\,Gyr), is dominated by the secondary RC stars, while the older RC population (hereafter RCold) is dominated by the primary RC stars with a mean age of about 4.6\,Gyr (covering age from 1\,Gyr to 10\,Gyr). Both RC populations are younger than the typical thick disc or the halo populations, but the difference in age is sufficiently large to distinguish the temporal effect in the structure of the thin disc.

We finally select 73,278 RCold stars and 32,423 RCyoung stars. Their distances are estimated according to Paper I, where the absolute magnitude of RC stars were fitted with isochrones and achieved a distance uncertainty of about 10\%. The spatial coordinates for these samples are converted to the Galactocentric cylindrical coordinates in which $R$ and $Z$ represent for the Galactocentric radius and the height above the Galactic mid-plane, respectively. We adopt the Sun position: $R_0=8$ and $Z_0=0.027$\,kpc (Chen et al.~\cite{chen2001}). The RC stars are mostly located from $150^\circ$ to $210^\circ$ in Galactic longitude and well cover the outer disc from $R\sim9$ to 14\,kpc.

Moreover, Liu et al.~(\cite{liu2017a}) developed a statistical method to derive the stellar density for a certain stellar population. According to this approach, the stellar densities have been determined separately for all the RC samples (i.e., RCold+RCyoung, hereafter RCall), the RCold, and RCyoung samples.

\section{Results}\label{sect:result}
\begin{table*}
\centering
\caption{The best-fit model parameters and their uncertainties for the vertical density profiles at various $R$ bins.}
\label{table:verticalmodel}
\begin{tabular}{c|c|c|c|c|c|c}
\hline\hline
$\rm{R(kpc)}$ \ & \ & \ 9.0 & \ 9.5 & \ 10.0  & \ 10.5 &\ 11.0  \\ 
\hline
$\rm{RC_{all}}$ & $\ln\nu(R)$ &  $-7.02_{-0.01}^{+0.02}$ & $ -7.53_{-0.02}^{+0.02}$ & $-7.87_{-0.01}^{+0.02}$ & $-8.12_{-0.02}^{+0.02}$ & $-8.31_{-0.02}^{+0.02}$  \\
 & $\rm{h_{Z,thin}(pc)}$ & $166.3_{-2.2}^{+2.3}$ & $211.7_{-3.4}^{+3.4}$ & $245.5_{-5.1}^{+5.1}$ & $270.8_{-6.3}^{+7.1}$ & $292.2_{-9.5}^{+9.8}$ \\
 & $\rm{h_{Z,thick}(pc)}$ & $641.7_{-1.2}^{+1.4}$ & $647.1_{-2.1}^{+3.3}$ & $651.9_{-4.2}^{+2.7}$ & $648.2_{-5.4}^{+5.3}$ & $650.1_{-8.2}^{+7.3}$ \\
 & $f_{t}$ & $0.032_{-0.003}^{+0.002}$ & $0.054_{-0.005}^{+0.005}$ & $0.083_{-0.009}^{+0.008}$ & $0.123_{-0.013}^{+0.014}$ & $0.137_{-0.019}^{+0.019}$ \\
\hline
$\rm{RC_{old}}$ & $\ln\nu(R)$ &  $-7.42_{-0.01}^{+0.02}$ & $-7.92_{-0.02}^{+0.01}$ & $-8.26_{-0.02}^{+0.02}$ & $-8.55_{-0.02}^{+0.02}$ & $-8.74_{-0.02}^{+0.02}$   \\
 & $\rm{h_{Z,thin}(pc)}$ & $176.3_{-2.5}^{+2.6}$ & $226.3_{-3.8}^{+3.9}$ & $259.1_{-6.2}^{+6.1}$ & $294.4_{-8.8}^{+9.1}$ & $326.2_{-13.5}^{+13.0}$ \\
 & $\rm{h_{Z,thick}(pc)}$ & $648.1_{-2.1}^{+3.4}$ & $650.2_{-2.9}^{+3.3}$ & $653.4_{-7.2}^{+4.1}$ & $641.5_{-6.4}^{+11.3}$ & $651.8_{-17.3}^{+12.2}$ \\
 & $f_{t}$ & $0.038_{-0.003}^{+0.003}$ & $0.063_{-0.006}^{+0.006}$ & $0.099_{-0.011}^{+0.012}$ & $0.139_{-0.019}^{+0.012}$ & $0.143_{-0.030}^{+0.028}$ \\
\hline
$\rm{RC_{young}}$ & $\ln\nu(R)$ &  $-8.11_{-0.01}^{+0.02}$ & $-8.64_{-0.02}^{+0.02}$ & $-8.99_{-0.02}^{+0.02}$ & $-9.18_{-0.02}^{+0.02}$ & $-9.34_{-0.02}^{+0.02}$   \\
  & $\rm{h_{Z,thin}(pc)}$ & $146.2_{-1.5}^{+1.6}$ & $183.9_{-2.2}^{+2.2}$ & $222.1_{-3.4}^{+3.6}$ & $236.3_{-4.4}^{+4.7}$ & $255.7_{-5.9}^{+6.2}$ \\
 & $\rm{h_{Z,thick}(pc)}$ & $655.7_{-4.3}^{+2.4}$ & $652.1_{-3.1}^{+2.8}$ & $649.7_{-4.3}^{+5.0}$ & $653.6_{-6.4}^{+4.9}$ & $650.9_{-8.7}^{+9.3}$ \\
 & $f_{t}$ & $0.015_{-0.001}^{+0.001}$ & $0.033_{-0.003}^{+0.002}$ & $0.045_{-0.005}^{+0.004}$ & $0.076_{-0.007}^{+0.007}$ & $0.081_{-0.009}^{+0.009}$ \\
\hline\hline
$\rm{R(kpc)}$ \ & \ & \ 11.5 &\ 12.0 &\ 12.5 &\ 13.0 &\ 13.5 \\ 
\hline
$\rm{RC_{all}}$ & $\ln\nu(R)$  & $-8.47_{-0.02}^{+0.02}$ & $-8.64_{-0.03}^{+0.02}$ & $-8.69_{-0.05}^{+0.06}$ & $-8.89_{-0.06}^{+0.07}$ & $-9.16_{-0.06}^{+0.07}$  \\
 & $\rm{h_{Z,thin}(pc)}$  & $303.1_{-11.8}^{+12.3}$ & $328.1_{-15.3}^{+15.9}$ & $315.7_{-23.2}^{+25.2}$ & $334.1_{-28.1}^{+31.2}$ & $424.2_{-40.3}^{+28.7}$\\
 & $\rm{h_{Z,thick}(pc)}$ & $648.6_{-9.2}^{+13.4}$ & $657.1_{-12.5}^{+10.1}$ & $654.9_{-20.0}^{+16.7}$ & $649.9_{-15.4}^{+25.3}$ & $655.2_{-28.2}^{+27.1}$ \\
 & $f_{t}$ & $0.175_{-0.022}^{+0.021}$ & $0.190_{-0.030}^{+0.026}$ & $0.193_{-0.035}^{+0.030}$ & $0.221_{-0.045}^{+0.038}$ & $0.132_{-0.088}^{+0.098}$ \\
\hline
$\rm{RC_{old}}$ & $\ln\nu(R)$ & $-8.93_{-0.03}^{+0.02}$ & $-9.12_{-0.03}^{+0.03}$ & $-9.21_{-0.06}^{+0.05}$ & $-9.36_{-0.06}^{+0.07}$ & $-9.62_{-0.06}^{+0.06}$  \\
 & $\rm{h_{Z,thin}(pc)}$  & $325.7_{-15.8}^{+16.5}$ & $366.3_{-21.9}^{+23.8}$ & $354.8_{-31.2}^{+34.7}$ & $346.5_{-32.6}^{+36.6}$ & $445.6_{-32.1}^{+23.5}$ \\
 & $\rm{h_{Z,thick}(pc)}$  & $649.0_{-21.2}^{+22.4}$ & $654.1_{-22.1}^{+20.7}$ & $652.2_{-28.2}^{+26.4}$ & $658.2_{-25.4}^{+28.9}$ & $649.0_{-28.0}^{+47.1}$ \\
 & $f_{t}$ & $0.218_{-0.035}^{+0.032}$ & $0.217_{-0.056}^{+0.046}$ & $0.211_{-0.069}^{+0.049}$ & $0.260_{-0.057}^{+0.045}$ & $0.101_{-0.070}^{+0.113}$ \\
\hline
$\rm{RC_{young}}$ & $\ln\nu(R)$ & $-9.46_{-0.02}^{+0.02}$ & $-9.59_{-0.03}^{+0.02}$ & $-9.65_{-0.05}^{+0.06}$ & $-9.83_{-0.07}^{+0.08}$ & $-10.12_{-0.07}^{+0.08}$ \\
 & $\rm{h_{Z,thin}(pc)}$ & $262.4_{-7.2}^{+7.3}$ & $286.1_{-8.9}^{+9.3}$ & $301.3_{-16.1}^{+17.2}$ & $289.2_{-19.2}^{+23.1}$ & $370.0_{-38.4}^{+38.9}$ \\
 & $\rm{h_{Z,thick}(pc)}$ & $644.7_{-7.2}^{+8.4}$ & $657.0_{-9.3}^{+10.9}$ & $650.8_{-17.2}^{+22.2}$ & $652.9_{-20.8}^{+21.3}$ & $654.1_{-39.1}^{+38.9}$ \\
 & $f_{t}$  & $0.100_{-0.011}^{+0.011}$ & $0.109_{-0.013}^{+0.013}$ & $0.107_{-0.022}^{+0.019}$ & $0.210_{-0.025}^{+0.024}$ & $0.174_{-0.081}^{+0.060}$ \\
 \hline\hline
\end{tabular}
\end{table*}

\begin{figure}
\centering
\includegraphics[width=1.0\textwidth,trim=30 0 0 0]{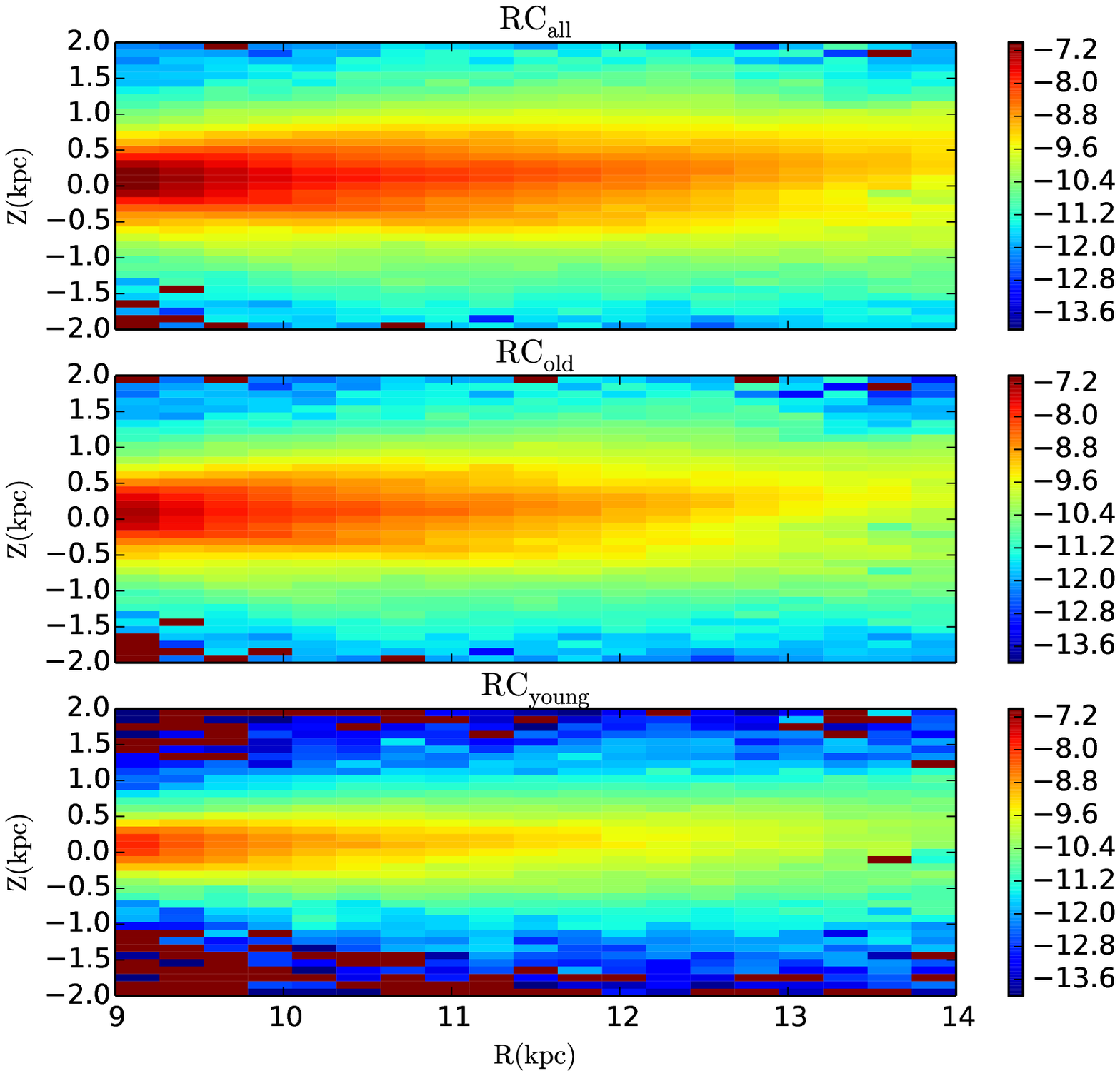}
\caption{The stellar density maps for the RCall (top panel), RCold (middle panel), and RCyoung (bottom panel) in the $R$--$Z$ plane. The colors indicate values of the $\ln\nu$.}
\label{fig:imfigure}
\end{figure}

Figure~\ref{fig:imfigure} presents the stellar density maps in the $R$--$Z$ plane for the RCall, RCold, and RCyoung populations from top to bottom, respectively. First, it clearly shows from the color coded $\ln\nu$ that the disc of the RCyoung population is substantially thinner than that of the RCold population. Second, the color coded logarithmic stellar densities indicate that the RCyoung is less dense than the RCold population. This means that the stellar outer disc is dominated by the older populations. The density map for the RCall is more similar with that for the RCold, confirming that the older population is indeed prominent.

To quantify the structures accounting for the complicated radial features in the outskirts, e.g. the flare, we split the stars into various $R$ bins and fit the 1-D vertical density profile along $Z$ in each $R$ bin. We assume that the disc is composed of two components and the density profile is separable about $R$ and $Z$ such that
\begin{equation}\label{eq:densitymodel1}
\nu(R,Z) = \nu(R|Z=0)\left(\nu_{thin}(Z|R)+\nu_{thick}(Z|R)\right),
\end{equation} 
where $\nu(R|Z=0)$ is the stellar density at $Z=0$. The two disc components in the right-hand side follows a hyperbolic secant squared profiles (Kruit~\cite{kruit1988}):
\begin{equation}\label{eq:sech1}
\nu_{thin}(Z|R)=\left(1-f_t(R)\right){\rm sech}^2\left(\frac{Z}{2h_{z,thin}(R)}\right)
\end{equation}
and
\begin{equation}\label{eq:sech2}
\nu_{thick}(Z|R)=f_t(R){\rm sech}^2\left(\frac{Z}{2h_{z,thick}(R)}\right),
\end{equation}
where $h_{z,thin}(R)$ and $h_{z,thick}(R)$ are the scale heights of the thin and thick components, respectively, and $f_t(R)$ is the fraction of the thick component. Because \del{the} RC stars are mostly metal-rich and they are selected with [Fe/H]$>-0.6$\,dex (Wan et al.~\cite{wan2015}), only very few halo stars should be included in the samples. Therefore, the halo component in the star count model is negligible.

\begin{figure}
\centering
\includegraphics[width=0.8\textwidth]{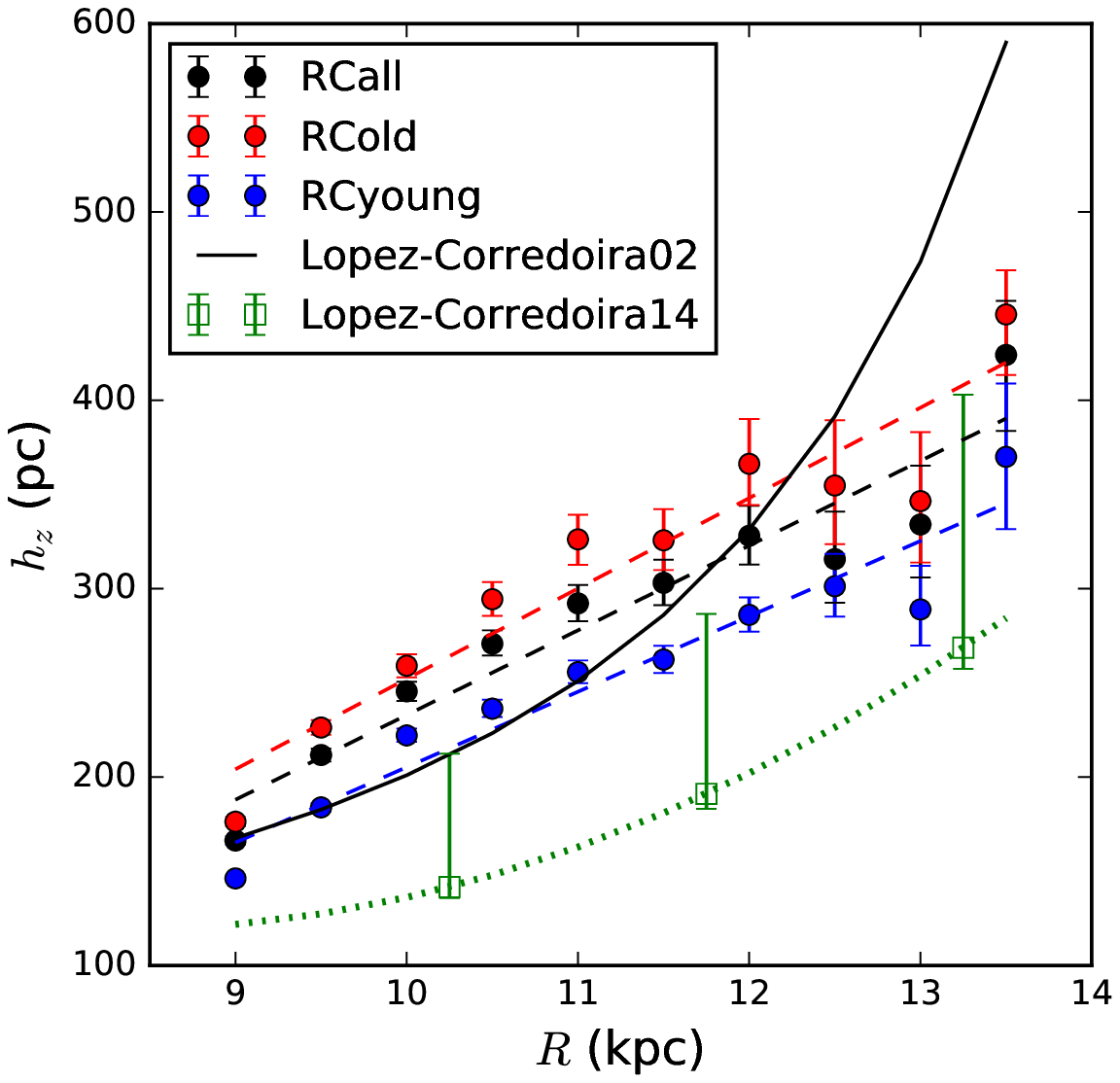}
\caption{The derived scale heights as functions of $R$ for the RCall (black dots), RCold (red dots), and RCyoung (blue dots) populations, respectively. The dashed lines represent the best linear fits about the scale heights for the corresponding populations with same colors. The black solid line is the scale height from L{\'o}pez-Corredoira M et al.~(\cite{lopez02}) divided by $2$. The green dotted line and three green squares with error bars indicates the scale height from L{\'o}pez-Corredoira \& Molg{\'o}~(\cite{lopez14}) divided by 2 as well.}\label{fig:scaleheight}
\end{figure}

The data is separated into 10 $R$ bins from 9 to 13.5\,kpc with the bin size of 0.5\,kpc. At each $R$ bin, we fit the disc model described by Eqs.~(\ref{eq:densitymodel1})-(\ref{eq:sech2}) with the Markov chain Monte Carlo technique\footnote{We use \emph{emcee} package (~\cite{foremanmackey2013}) to run the Markov chain Monte Carlo simulation.} for the RCall, RCold, and RCyoung, respectively. The best-fit model parameters with uncertainties are listed in Table~\ref{table:verticalmodel}.

Note that the thick disc population may not be completely sampled since the cut of [Fe/H]$>-0.6$\,dex removes lots of metal-poor thick disc stars. Therefore, in this work, we only focus on the thin disc components. 

The best-fit scale heights of the thin disc component for the RCall, RCyoung and RCold are shown in Figure~\ref{fig:scaleheight} with black, blue, and red dots, respectively. Over all $R$ bins, the scale heights for the RCyoung are smaller than those for the RCold population by a factor of 82\%. In other word, the RCyoung is consistently thinner than the RCold in the range of $R$ from 9 to 13.5\,kpc. This is expected if the thin disc is heated by giant molecular clouds. In such a scenario, the older population has been heated for longer time and hence thicker than the younger one. 

\begin{figure}
\centering
\includegraphics[width=0.8\textwidth]{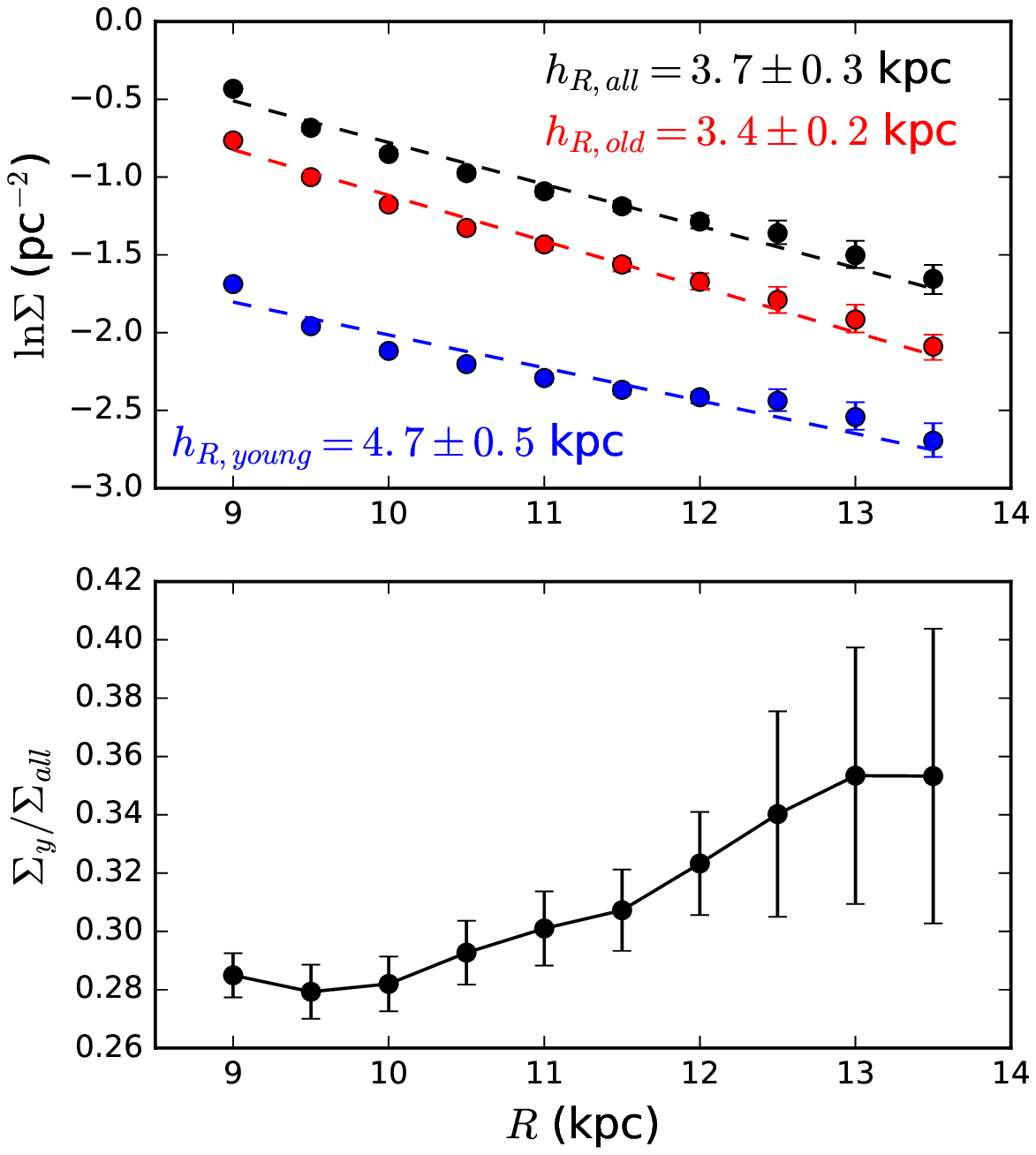}
\caption{The top panel displays the surface stellar densities, in logarithmic form, as functions of $R$ for the RCall (black dots), RCold (red dots), and RCyoung (blue dots) populations, respectively. The dashed lines represent for the best fit exponential profiles for the corresponding populations with the same colors. The bottom panel displays the ratio of the surface density for the RCyoung to RCall population.}
\label{fig:surfacedensity}
\end{figure}

It is also clearly evident that the flare is found in both RCyoung and RCold populations. Indeed, the scale heights increase from $\sim150$\,pc at $R=9$\,kpc to about $400$\,pc at $R=13.5$\,kpc for both populations. Although the increasing $h_z$ shows some oscillations, the flare can be empirically fitted with a linear model within the detecting range of $R$. The dashed lines shown in the figure represent for the best linear fits about $h_z$ as a function of $R$ for the populations with the same colors. The radial gradient of the scale heights are $45\pm5$, $48\pm6$, and $40\pm4$\,pc/kpc for the RCall, RCold, and RCyoung populations, respectively. The gradients for the RCyoung and RCold populations are not significantly different, implying that the intensity of the flaring seems not tightly related to the age of the thin disc populations. 

We then compare the flare from our result with L{\'o}pez-Corredoira et al.~(\cite{lopez02}), who modeled the flaring of the disc with an exponential function of $R$. Note that L{\'o}pez-Corredoira et al.~(\cite{lopez02}) adopted an exponentially declining vertical profile, while we apply a ${\rm sech}^2$ form. The scale height of the exponential profile is larger than that of the ${\rm sech}^2$ profile by about a factor of 2. Therefore, we divide the scale height model of L{\'o}pez-Corredoira et al.~(\cite{lopez02}) by 2 so that they can be compared with our result. The black solid line in the figure indicates the rescaled flaring scale height from L{\'o}pez-Corredoira et al.~(\cite{lopez02}). It is quite similar with the scale heights from this work at $R<11$\,kpc. However, at $R>11$\,kpc, the flaring scale heights modeled by L{\'o}pez-Corredoira et al.~(\cite{lopez02}) are significantly larger than our result. The discrepancy at $R>11$\,kpc may be caused by the different areas of the sky covered by each sample, or the different methods used to select RC stars. L{\'o}pez-Corredoira et al.~(\cite{lopez02}) selected RC stars from infrared photometry, while we select them from spectroscopic data. 

L{\'o}pez-Corredoira \& Molg{\'o}~(\cite{lopez14}) quantified the flare with a quadratic function of $R$, which is shown as green dotted line and three green squares with error bars to indicate the uncertainties in Figure~\ref{fig:scaleheight} (It is also divided by 2 to align with our sech$^2$ scale height). Although the gradient of the scale height from L{\'o}pez-Corredoira \& Molg{\'o}~(\cite{lopez14}) is quite similar with ours, the values of scale height are smaller than ours. Note that L{\'o}pez-Corredoira \& Molg{\'o}~(\cite{lopez14}) used F/G type dwarfs as tracers, which are younger than our samples. Normally, younger population has smaller scale height. Also, considering the larger uncertainties, their scale heights are only about 1-$\sigma$ lower than our result, which is not statistically significant.

On the other hand, the surface stellar density, $\Sigma(R)$, can be derived by integrating the vertical profile over $Z$. For the ${\rm sech}^2$ profile, it is easy to obtain that 
\begin{equation}\label{eq:Sigma}
\Sigma(R)=4\left((1-f_t(R))h_{z,thin}(R)+f_t(R)h_{z,thick}(R)\right)\nu(R|Z=0).
\end{equation}
The top panel of Figure~\ref{fig:surfacedensity} shows the stellar surface densities as functions of $R$ for the populations of RCall (black dots), RCold (red dots), and RCyoung (blue dots), respectively. The dashed lines indicate the best fit exponential profiles of $\Sigma$ for the corresponding populations with the same colors. The derived scale lengths are $3.7\pm0.3$, $3.4\pm0.2$, and $4.7\pm0.5$\,kpc for the RCall, RCold, and RCyoung, respectively. Note that the RCold population has larger scale height with smaller scale length, while the RCyoung population has smaller scale height with larger scale length. The trend that the younger populations have larger scale length and smaller scale height is qualitatively consistent with Bovy et al.~(\cite{bovy2012}).

The bottom panel of Figure~\ref{fig:surfacedensity} shows the ratio of the surface density of RCyoung to RCall population, $\Sigma_y/\Sigma_{all}$. It is around $28$\% at $R<10$\,kpc, while, at $R\sim13$\,kpc, the $\Sigma_y/\Sigma_{all}$ moderately increases to $35$\% with relatively larger uncertainty. This means that the relative number of the younger stars becomes slightly larger in the outer disc, although the older population is still dominant in the outskirt region. 

Note that Martig et al.~(\cite{martig2016}) claimed that the median age of the RC stars located at $0<|Z|<0.5$\,kpc decreases from about 8\,Gyr at $R\sim5$\,kpc to 4\,Gyr at $\sim8$\,kpc. From $R=8$ to $\sim14$\,kpc, the median age is roughly flat or moderately declines (see their figures 1 and 3). By separating the stars into two populations with different ages, we show that the age distribution of the stellar populations may be variable at $9<R<13.5$\,kpc in the sense that the fraction of the younger stars slightly increases. However, this may not substantially leverage the median age of the whole outer disc, since the older population occupies more than 65\% in this regime.
 
\begin{figure}
\centering
\includegraphics[width=1.0\textwidth]{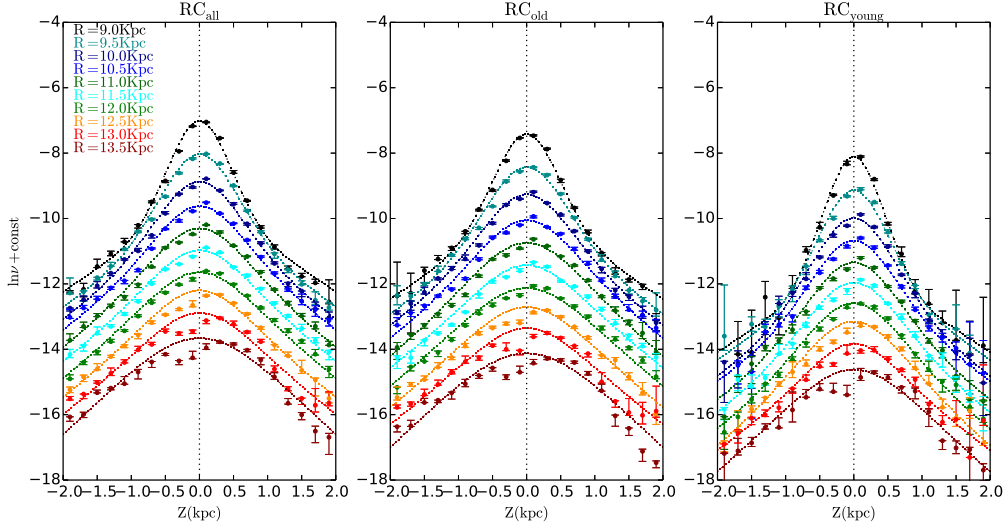}
\caption{The vertical density profiles for the RCall, RCold, and RCyoung populations (from left panel to right respectively) in different $R$ bins. The colors code the Galactocentric radii. The dash lines stand for the models described by Eqs.~(\ref{eq:densitymodel1})-(\ref{eq:sech2}) corresponding to the data with the same colors. In order to avoid overlapping, the vertical stellar density profiles are tiled with different offsets in $\ln\nu$ for different $R$ bins.}
\label{fig:verticaldensity}
\end{figure}

\section{Discussions \& Conclusions}\label{sect:conclusion}
In general, the star counts along the Galactic disc are often suffered from the interstellar extinction. Although the extinction is not severe in the Galactic anti-centre direction, it may still affect the stellar density determinations. Figure~\ref{fig:imfigure} shows an elongated shallow dip located between $R=11$ and $14$\,kpc and $Z\sim-0.3$\,kpc. It is very likely due to the incompleteness of the photometric data induced by the extinction. Indeed, Green et al.~(\cite{green2015}) indicated that there are clouds with larger interstellar extinction located at distance of 794--1995\,pc, $l\sim180^\circ$, and below $b=0^\circ$, which may be associated with the shallow dip.

Although the interstellar extinction may affect the stellar density measurements in some regions of the $R$--$Z$ plane, especially in the south disc (south disc presents Z$>0$, thereafter, north disc presents Z$<0$), at Z close to 0, it does not play an important role in the overall shape of the vertical density profiles. Figure~\ref{fig:verticaldensity} shows that the vertical profiles for most of the $R$ bins are well fitted with the models, except the lowest two profiles, which corresponded to the two largest $R$ bins. Even excluding the last two $R$ bins, Fig.~\ref{fig:scaleheight} shows that the linearly increasing profiles are still substantial. Therefore, the extinction does not change our result.

With the data of HI region in the outer Galactic disc, Levine, Blitz \& Heiles~(\cite{levine06}) found the asymmetry in the warp (especially at galactocentric azimuth $90^\circ$ and $-90^\circ$, see Fig.6, Fig.7, Fig.8 in their papar), which is claimed as prominent feature. However, in the anti-centre direction (which galactocentric azimuth is around $0^\circ$), the height of gas is almost $0$kpc at different galactocentric radius (see Fig.6, Fig.7, Fig.8 in their papar). L\'{o}pez-Corredoira et al.~(\cite{lopez02}) also claimed that warped and unwrapped models provide equivalent fits to the data towards the anti-centre direction (at longitude $150^\circ$ to $210^\circ$, see their figure 14). Moreover, LAMOST survey covers a larger area in the north disc than the south at longitudes $180^\circ$ to $210^\circ$. On the other hand, at longitudes $150^\circ$ to $180^\circ$, LAMOST survey covers a larger area in the south disc than the north. Hence the footprints of LAMOST survey not follow the warp of the Galactic disc, which means that the stellar vertical density distribution derived from LAMOST data should not be impacted by the Galactic warp. In other words, it is hard to exhibit the geometry of the warp from LAMOST data in the anti-centre direction. Therefore, the dip shown at 11$<R<$14\,kpc is not likely due to the asymmetry of the warp.

From the stellar density profiles, we find that the younger and older RC populations in the outer disc are significantly flared. The flaring disc traced by both populations show increasing scale heights with the Galactocentric radius of $R$. The intensity of flaring seems not substantially correlated with the age of the thin disc populations. However, our RCyoung and RCold populations have age overlaps from 2\,Gyr to 4\,Gyr (see section.\ref{sect:data} and Fig.3 in the Tian et al.~(\cite{tian16})). This age overlaps may cause that the intensity of flaring for our two populations are likely the same. On the other hand, the age of our most of RC stars are younger than 6\,Gyr, which may not be old enough to distinguish the difference of flaring with different populations. 

The scale length of the surface density for the RCyoung population is $4.7\pm0.5$\,kpc, which is significantly larger than the scale length of $3.4\pm0.2$\,kpc for the RCold population. Meanwhile, the scale heights for the RCyoung population are systematically smaller than those for the RCold population. Moreover, the fraction of the younger population is determined from the surface stellar density of the younger and older populations. It shows that the fraction of the RCyoung population moderately but consistently increases from $28$\% at $R=9$\,kpc to $35$\% at $R\sim13$\,kpc.

Liu et al.~(\cite{liu2017a}) claimed that the scale length of thin disk is $\sim 1.6\pm0.1$ by using RGB stars from LAMOST data. It is quite smaller than our results. This is possibly because that, in the mean, RGB stars are older than RC stars. More specifically, younger RGB stars are more massive than old RGB stars. A star with 2 $\rm M_{\sun}$ at solar metallicity can only stay for about 44\,Myr in RGB stage, but can stay for more than 100\,Myr in the stage of Helium-core burning. As a comparison, a star with 1 $\rm M_{\sun}$ can stay for more than 900\,Myr in the stage of hydrogen-shell burning and more than 100\,Myr in the helium-core burning phase. Hence, the young and massive stars have much less probability to be sampled in the RGB stage than in the RC stage (see Table 5-2, Binney\&Merrifield~(\cite{binney98})). Therefore, RGB stars and RC stars are two different populations, which are with different scale length. Moreover, the RC stars used in this work have [Fe/H]$>-0.6$ dex, but in Liu et al.~(\cite{liu2017a}), RGB stars with all range of [Fe/H] are used to derived the surface stellar density profile. Although the age-metallicity relation is quite flat, metal-poor stars tend to be older than metal-young ones. Therefore, averagely, the RGB stars are older than RC stars, which may have a smaller scale length.

Amores et al.~(\cite{amores17}) identified a dependence of the thin disk scale length with age by using 2MASS data and they found that the value of scale length is in the range from $\sim$3.8 kpc (for youngest stars ($<1$Gyr) in their sample) to $\sim$2.0 kpc (for oldest stars ($\sim 8.5$Gyr) in their sample). Their value of scale length for youngest stars is similar to our results, but for oldest stars, the value is smaller than ours. The reason may be that their youngest samples have similar age to ours and the oldest samples are older than ours.

It is reasonable to check our results with the RC samples cut at other ages. Unfortunately, our isochrones-based separation lines for RC populations (see Section~\ref{sect:data}) can not be changed, because it is essentially determined by the helium-flash mass.  If we change our separation lines, there would be more younger RC stars to contaminate the RCold population or more older RC stars to contaminate the RCyoung population (see Fig.2 in Tian et al.~(\cite{tian16})). In the other words, our separation lines can only be at the age $\sim 2$Gyr.  However, combined with other stellar tracers (such as dwarfs, turn-off stars),  our samples could have a larger age range, which may more appropriate for study the dependence of disc structure with different populations. We will check this in the future.

\section*{Acknowledgements}
This work is supported by the National Key Basic Research Program of China 2014CB845700. CL acknowledges the National Natural Science Foundation of China under grants 11373032 and 11333003. Guoshoujing Telescope (the Large Sky Area Multi-Object Fiber Spectroscopic Telescope LAMOST) is a National Major Scientific Project built by the Chinese Academy of Sciences. Funding for the project has been provided by the National Development and Reform Commission. LAMOST is operated and managed by the National Astronomical Observatories, Chinese Academy of Sciences.

\clearpage
\end{document}